\documentclass[twocolumn,superscriptaddress,prl,floatfix]{revtex4}
\usepackage{amsfonts,amsmath,amssymb}
\usepackage{graphicx}
\begin{document}
\title{How human drivers control their vehicle}
%Driving your car -- towards a modelling of human car following behaviour
\author{Peter Wagner}
\affiliation{Institute of Transport Research, German Aerospace Center
  (DLR), Rutherfordstrasse 2, 12489 Berlin, Germany.}
\date{\today }
\begin{abstract}
The data presented here show that human drivers apply a discrete noisy control mechanism to drive their vehicle. A car-following model built on these observations, together with some physical limitations (crash-freeness, acceleration), led to non-Gaussian probability distributions in the speed difference and distance which are in good agreement with empirical data. 
All model parameters have a clear physical meaning and can be measured. Despite its apparent complexity, this model is simple to understand and might serve as a starting point to develop even quantitatively correct models.
\end{abstract}

\maketitle

\section{Introduction}

Modelling the process by which a driver controls her vehicle has been done since 1950.
% \cite{Reuschel}. 
So far, no commonly agreed model has been emerged. Even worse, the advent of the cellular automaton models (see \cite{Chowdhury} for a review) has sparked a burst of new models trying to describe (at least) the car following process of one car driving behind another one.

Most driving models assume (for reviews see \cite{Helbing,OR-review,Brackstone-review}) an instantaneous or even delayed reaction of the driver to the surrounding situation, i.e.\ the driving law can be formulated as a stochastic differential equation (SDE):
\begin{equation}
\dot v = A(g,v,V) + D(g,v,V)\xi .
\label{eq:genODE}
\end{equation}
Here, $g$ is the headway to the vehicle in front (distance from front bumper of the following vehicle to the head bumper of the lead vehicle), $v$ is the speed of the following car, $V$ the speed of the leading car, $\xi$ is a noise term which is restricted in size (acceleration is limited, and so is the noise), and $A(\cdot)$ and $D(\cdot)$ are two functions describing the reaction of the human driver to the situation in front of her vehicle.

Obviously, an equation like this one ignores two important features of human driving and of human actions in general. Firstly,  humans usually plan ahead, and secondly, the type of control humans apply is not continuous, but discrete in time: they act only at certain moments in time. These specific moments have been named action-points \cite{Todosiev1963,Michaels1963}, a name that will be used in the following.

\begin{figure}[h]
\begin{center}
\includegraphics[width=0.8\linewidth]{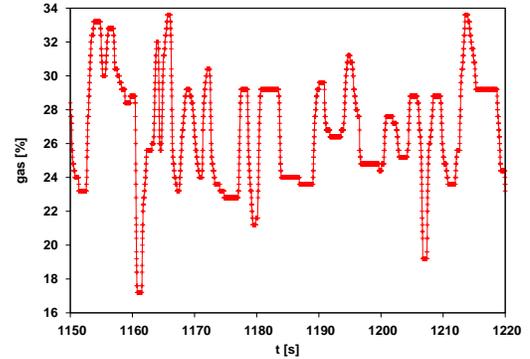}
\caption{The amount of gas (in \% ranging from 0 to 100) applied by the driver during a normal drive on a rural road. In 95 \% of all data-points in this time-series the amount of gas between subsequent data-points did not change.}
\label{fig:viewcar}
\end{center}
\end{figure}

\begin{figure}[h]
\begin{center}
\includegraphics[width=0.8\linewidth]{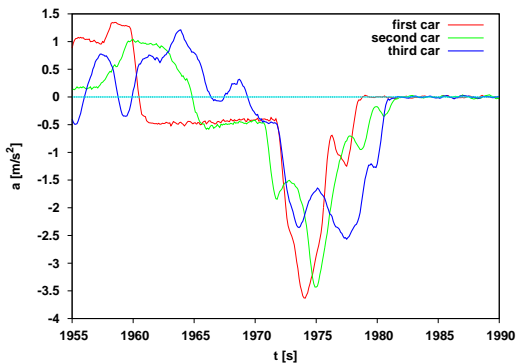}
\caption{Acceleration time-series of the leading (1st) and two following vehicles (2nd and 3rd). Usually, changes in acceleration occur time delayed, while around time $t=1970$ this behaviour changes: the 2nd vehicle brakes before the 1st one, while the 3rd one brakes in synchrony with the 1st car.}
\label{fig:earlyBrake}
\end{center}
\end{figure}

While the second assumption can be proven by analysing data from car-following experiments, see Fig.~\ref{fig:viewcar}, the first assumption has to be classified as a conjecture. Strictly speaking, there is no way of actually observing this planning process. The best that can be done is to find examples in data for such a planning. One interesting place where this can be observed is a courtesy lane change at freeway entrances; another one is the advance braking in front of a red traffic light. In Fig.~\ref{fig:earlyBrake}, an example for the latter behaviour is presented. Here, the acceleration times series of three vehicles are shown, with the following vehicles reacting before or in synchrony with the lead car. Of course, both types of behaviours contradict the assumption of an instantaneous driving law, i.e.\ Eq.~(\ref{eq:genODE}).

The planning ahead may be described by a trajectory computed in advance in the phase space $(x(t),v(t),a(t))$ of the following vehicle ($x(t)$ being a position along the road, in general it involves the lateral dynamics as welln), i.e.\ including a planned velocity and acceleration time course. First steps to formulate and analyse such models have been put forward in \cite{BRDM,RDM}, here a more specific example will be added.

\section{The data used}

Mainly two types of data have been used to support the results in this article. The first are several data-sets from an instrumented vehicle, where speed, distance to the leading vehicles, acceleration, the amount of gas, viewing angle, steering angle and many more data have been recorded with a high temporal resolution during several drives under normal to heavy traffic conditions. Different subjects drove the car; they were fully informed about the experiments.

The second data-set is from the NGSIM project \cite{NGSIM} and consists of several thousand trajectories of vehicles driving along two California freeways. The data have been recorded by video cameras, therefore only the positions of the cars (recorded in 0.1 s resolution) are the primary data, anything else like speeds has been computed from the trajectories.

The data consist of vehicle trajectories. One may wonder about how generic such a trajectory is, and here the assumption is made, that any trajectory is the reaction of a particular driver to a particular environment, and that this reaction can be parameterized if a suitable model of the car driving process has been found. Nevertheless, the approach taken here focuses on more robust features, that is, on probability distributions of various observables instead of the observables themselves.

The most prominent distributions to study are the ones for the acceleration $p(a)$, the speed differences between two vehicles $p(\Delta v)$, the headway distribution $p(T)$, where $T = g/v$ is the scaled distance between the vehicles, and the compound distribution $p(\Delta v,T)$, which is a sensitive measure of the interaction between two vehicles.

\section{A simple model}

\begin{figure}[h]
\begin{center}
\includegraphics[width=0.7\linewidth]{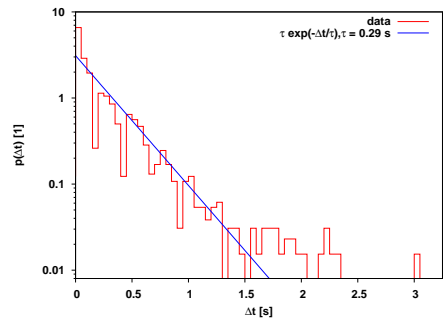}
\caption{The distribution of the time intervals between subsequent action points. The exponential function is a fit to the data, with $\tau = 0.29(5) $ s. The time intervals with $\Delta t \ge 2$ s stem from episodes where the car has been stopped, for instance by a red light.}
\label{fig:ap-head}
\end{center}
\end{figure}

It has been argued above that the control process applied by humans is discrete and noisy. The discreteness has been demonstrated in Fig.~\ref{fig:viewcar}. The randomness can be seen in  Fig.~\ref{fig:ap-head}, where the distribution $p(\Delta t)$ of time intervals $\Delta t$ between subsequent action-points is displayed.
This distribution follows an exponential distribution quite close, which can be understood as a simple process: in any instant of time the driver decides randomly whether he should change acceleration or not. Of course, there might be more sensible reasons to change acceleration, but mainly the action-points happen more or less without reasons, i.e.\ randomly. 

\begin{figure}[h]
\begin{center}
\includegraphics[width=0.7\linewidth]{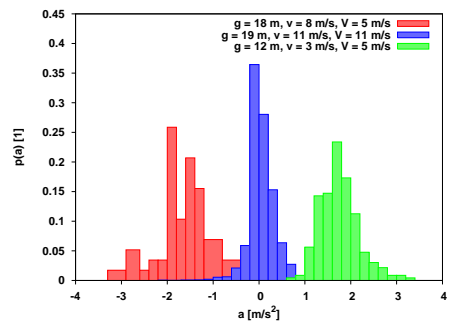}
\caption{The distribution of accelerations for negative speed difference (approaching lead vehicle), zero speed difference, and positive speed difference. Data are from ten cars on a Japanese test track \protect{\cite{DGPS-RTK}}, with relatively small speeds. All distributions are sampled from a small phase space box of size 1 m $\times$ 1 m/s $\times$ 1 m/s and are centered on the values indicated in the legend.}
\label{fig:probA}
\end{center}
\end{figure}

Additionally, the randomness is not only in time, the acceleration itself is not a very well defined function of distance and speed-difference. This can be seen from sampling acceleration values from a small phase-space interval $(g \pm \delta g/2, v \pm \delta v/2, V \pm \delta V/2)$. Three resulting distributions $p_{g,v,V} (a) \delta g \, \delta v \, \delta V$ are displayed in Fig.~\ref{fig:probA}. Typically, those distribution have standard deviation around 0.4 m/s$^2$, which will be interpreted in the following as the acceleration noise.

The results obtained so far therefore motivate the following model. At the time $t_n$ of the $n$-th action-point the driver chooses a new acceleration value $a_n$ according to:
\begin{equation}
a_n = a_{\text{opt}} - \varepsilon_a \xi
\end{equation}
Here, $a_{\text{opt}} $ is an optimal acceleration value to be specified below, $\varepsilon_a$ is the size of the acceleration noise and $\xi$ is a random number in the interval $[0,1]$.

After that, the driver keeps acceleration constant (more precisely: the amount of gas, acceleration might still change due to changing air resistance or efficiency changes in the engine), so the dynamics until the next action-points occurs follows simple laws:
\begin{eqnarray*}
v(t) & = & v_n + a_n \, (t - t_n) \\
x(t) & = & x_n + v_n \, (t - t_n) + \frac{1}{2} a_n \, (t - t_n)^2 \quad t \in [t_n,t_{n+1}]
\end{eqnarray*}
Here, $x_n, v_n$ are the values of the variables $x$ and $v$ at the time of the $n$-th action-point. Note, that this equation is mathematically a map. This may explain why even simple time-discrete models (which are maps with regular time intervals instead of the random time intervals here) can model traffic flow quite successfully.

As demonstrated already, the time intervals $\Delta t = t_{n+1} - t_n$ between the action-points are exponentially distributed, which can be translated into a time-discrete set-up as a certain probability $p_{\text{AP}}$ that an action-point will occur in a given time-step. The action-points itself are selected depending either on a random number drawn in any time-step with $\xi < p_{\text{AP}}$, or if the optimal acceleration is smaller than the current acceleration $a$ minus $\varepsilon_a$.

The optimal acceleration $a_{\text{opt}}$ can be computed by modeling the planning process of the driver. To drive safely during the short-term planning horizon $\tau$, it should be possible to safely stop the vehicle behind the leading vehicle later on. That means, that a driver chooses the maximum acceleration $a_{\text{opt}}$ which fulfils the following condition:
\begin{equation}
d(v + a_{\text{opt}}  \, \tau) + v \, \tau + \frac{1}{2} a_{\text{opt}} \, \tau^2 \le d(V) + g 
\label{eq:safety}
\end{equation} 
Here, $d(\cdot)$ are the braking distances, by assuming a constant (comfortable) deceleration $b$ which is the same for both drivers, this equation can be solved to yield:
\begin{equation}
a_{\text{opt}} = - \frac{v}{\tau} - \frac{b}{2} + \sqrt{\left (\frac{v}{\tau} - \frac{b}{2} \right )^2 + \frac{2\,b\,g + V^2 - v^2}{\tau^2} }
\label{eq:asafe}
\end{equation} 
This expression must be limited to a maximal acceleration: for a realistic model, $a_{\text{opt}} \le a_{\text{max}}(1-v/v_{\text{max}})$ has to be enforced. 

Taken anything together, this model has just seven parameters: the physical limitations car-length $\ell$, maximum speed $v_{\text{max}}$, and maximum acceleration $a_{\text{max}}$, and the behavioural parameters preferred deceleration $b$, acceleration noise $\varepsilon_a$, the action point probability $p_{\text{AP}}$, and the minimum preferred headway distance $\tau$. In principle, the maximal possible deceleration is another physical parameter, fortunately the decelerations of the model never reach unphysical deceleration values. 

This concludes the definition of the model. Note, that despite the rather complicated look of Eq.~(\ref{eq:asafe}), its geometric appearance is almost linear. Therefore it is very likely, that humans are capable of learning at least a certain approximation to this function, no reason to do fairly complicated math while driving.

\section{Running the simulations}

To compare the model with the data, simulations with $N=100$ vehicles have been run, either in a loop or by following a lead vehicle driving with constant speed. The time-step size has been set to $h=0.2$ s. Simulations with a smaller time-step size yield the same results, which is to be expected since the dynamical equations are the exact solutions of the model. To use more than one vehicle following a lead vehicle is important, since the behaviour at the end of a platoon differs from the behaviour directly behind the lead car. 

The following Figures show the results of the simulation compared to the real data. First, the headway distributions $p(T)$ are compared. In this case, as indicated by the semi-logarithmic plot, the distributions (simulation and data) follow quite closely a gamma distribution ($p(T) \propto T^\gamma \exp(-T/m)$). This is in agreement with the standard assumption \cite{Cowan:1976,Luttinen:1992}. Note however, that under some circumstances this distribution may change into a different form which can be obtained by the transformation $T \to 1/T$. This will be detailed elsewhere.
\begin{figure}[h]
\begin{center}
\includegraphics[width=0.8\linewidth]{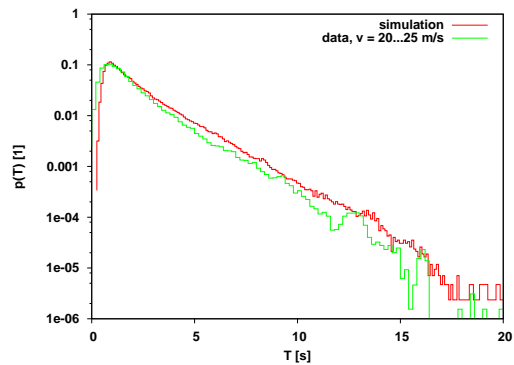}
\caption{The distribution of the time headways $p(T)$; comparison between simulation and data.  The parameters chosen are: $p_{\text{AP}}$=0.2, $\varepsilon_a = 0.4$ m/s$^2$, $v_{\text{max}}=30$ m/s, $a_{\text{max}}=2$ m/s$^2$, $b = 0.8$ m/s$^2$, $\tau=0.1\ldots0.5$ s, and $\ell=5.5$ m.}
\label{fig:pT}
\end{center}
\end{figure}
Secondly, the distribution of the speed differences is shown in Fig.~\ref{fig:pDV}. Again, good agreement between simulation and reality could be seen.
\begin{figure}[h]
\begin{center}
\includegraphics[width=0.8\linewidth]{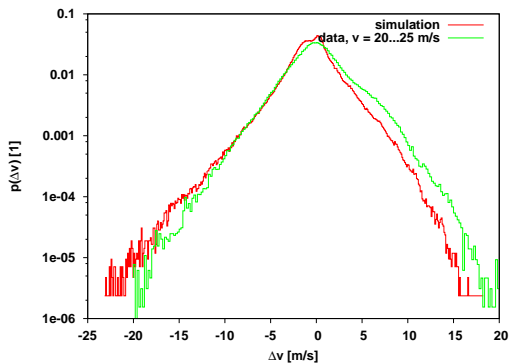}
\caption{The distribution of the speed differences $p(\Delta v)$. The simulation parameters are as in \protect{Fig.~\ref{fig:pT}}. }
\label{fig:pDV}
\end{center}
\end{figure}
%
%\section{A simplified theory}
The following simple consideration aids in understanding $p(\Delta v)$. Since the action-point dynamics makes it very hard for the driver to arrive at a fixed point of the car following dynamics $a=0, \Delta v=0, g = g^\star$, the acceleration may be modelled by a simple SDE. By ignoring the relatively weak dynamical component in the $g$-direction and concentrating on the strong dynamics in $\Delta v$-direction a 1D equation can be assumed:
\begin{equation}
\dot{\Delta v} =  a \tanh(\alpha \Delta v) +  \varepsilon \xi
\end{equation}
The force on the right hand side is symmetric with respect to $\Delta v=0$, which seems a good approximation even for large $\Delta v$, see again Fig.~\ref{fig:pDV}. The somewhat surprising $\tanh(\cdot)$--nonlinearity stem from the idea, that the driver mainly apply two acceleration values: positive ones for positive $\Delta v$, and negative ones for negative $\Delta v$. Obviously, this is a gross oversimplification, which can be justified by the result below. 

The Fokker-Planck equation for the stationary phase space density $w(\Delta v)$ to this Langevin equation can be solved exactly:
\begin{equation}
w(\Delta v) \propto \frac{1}{\varepsilon} \cosh \left ( \alpha \Delta v) \right )^{-\frac{2 a}{\alpha \varepsilon}} 
% \to \exp (-|\Delta v|) \quad \mbox{for $\Delta v \t \infty$}
\end{equation}
This is what is observed in Fig.~\ref{fig:pDV}, therefore some understanding of the origin of the $p(\Delta v)$--distribution is gained. 

\section{Summary and conclusions}

The model presented here has some limitations. For instance, it is not completely platoon stable, i.e.\ in a long platoon disturbances can amplify and finally lead to standing vehicles. Some details of the acceleration distribution (not shown) are not in full agreement to the data, and the restriction to two car interactions must be lifted, the NGSIM data for instance are from a six-lane freeway. Another limitation is that the model described here only models the operational driving process covering the next two seconds or so, while nothing is done to model tactical driving covering the next 10 seconds. 

Despite these limitations, it is capable to model human driving faithfully. It is important to recognize that the interaction between vehicles, together with the action-point dynamics, lead to the exponential distribution of the speed differences. Albeit this particular form of the distribution signals that $\Delta v = 0$ is a special value, the action-points hinder the formation of a stable fixed point of the car following process. Many models described in the literature assume such a fixed point; however most of them can be made more realistic by adding the action-point mechanism as described above. 

More facts have been learned about the interaction between cars: first of all, the interaction is controlled to a large part by $\Delta v$, the distances seem to be rather unimportant to the driver as long as they are in a certain comfortable range. This is similar to the model in \cite{BK-book}, but in contradiction to the so called optimal velocity models which model the interaction as a function of distance only. Interestingly, when a vehicle is in following mode the decelerations applied are rather small, typically the drivers control their vehicle not by applying the brakes but simply by stepping off the gas. This makes the distribution of $p(\Delta v)$ so amazingly symmetric; when switching to a larger $b$ in the model above, the (simulated) distribution becomes asymmetric.

Let us finally speculate about why humans drive in this manner: because it is simply much more comfortable to mince around a preferred distance than to actually fix it completely.

\section*{Acknowledgments}

Many thanks to T.~Nakatsuji and his Hokkaido group for sharing their data. The NGSIM project provided the beautiful trajectory data-sets, which for sure will help to advance traffic flow research. Data of the equipped car have been provided by J\"urgen Rataj, other data came from the group of Michael Schreckenberg, which are acknowledged here as well.


\begin{thebibliography}{99}
\bibitem{Chowdhury} D.~Chowdhury, L.~Santen, and A.~Schadschneider,
  Phys. Rep. \textbf{329}, 199 (2000).
%
\bibitem{Helbing} D.~Helbing, Rev.~Mod.~Phys. \textbf{73}, 1067
  (2001).
%
\bibitem{OR-review} K.~Nagel, P.~Wagner, and R.~Woesler, Oper. Res.
  {\bf 51} 681 (2003).

\bibitem{Todosiev1963} E.~P.~Todosiev, and L.~C.~Barbosa, Traffic
  Engineering {\bf 34}, 17 (1963/64).

\bibitem{Michaels1963} R.~M.~Michaels, Proceedings of the second
  international symposium on the theory of road traffic flow, 44 -- 59,
  OECD (1963).

\bibitem{DGPS-RTK} G.~S.~Gurusinghe, T.~Nakatsuji, Y.~Azuta,
  P.~Ranjitkar, and Y.~Tanaboriboon, Transp. Res. Rec. {\bf 1802}, 166 (2003).

\bibitem{Cowan:1976} R.~J.~Cowan, Transp.~Res. {\bf 9(6)}, 371 (1976).

\bibitem{Luttinen:1992} T.~Luttinen, Transp.~ Res.~Rec., {\bf 1365}, 111 (1992).

\bibitem{Brackstone-review} M.~Brackstone and M.~McDonald, 
Transp.~Res.~F, {\bf 2}, 181 - 196 (2000). 

\bibitem{BRDM} I.~Lubashevsky, P.~Wagner, and R.~Mahnke, Europ. Phys.
  J. B {\bf 32} 243 -- 247 (2003).

\bibitem{RDM} I.~Lubashevsky, P.~Wagner, and R.~Mahnke, Phys.~Rev.~E
  {\bf 68} 056109 (2003).

\bibitem{BK-book} B.~S.~Kerner, The Physics of Traffic, Springer,
  Berlin, Heidelberg, New York, 2004.

\bibitem{NGSIM} Next Generation Simulation Programme, http://ngsim.camsys.com/, accessed January 2006.

\end{thebibliography}
\end{document}